\documentclass[12pt, preprint]{aastex}
\usepackage{epsfig}
\usepackage{amsmath}
\usepackage{graphicx}

\makeatletter
\usepackage{fancyhdr}
\makeatother
\shorttitle{Plate Tectonics on Super-Earths}
\shortauthors{Valencia et al.}

\begin{document}

\title{Inevitability of Plate Tectonics on Super-Earths}

\author{Diana Valencia\altaffilmark{1}}
\affil{Earth and Planetary Sciences Dept.,
Harvard University, 20 Oxford Street, Cambridge, MA 02138}
\email{valenc@fas.harvard.edu}

\author{Richard J. O'Connell}
\affil{Earth and Planetary Sciences Dept.,
Harvard University, 20 Oxford Street, Cambridge, MA 02138}
\email{oconnell@geophysics.harvard.edu}

\and
\author{Dimitar D. Sasselov}
\affil{Harvard-Smithsonian Center for Astrophysics,
60 Garden Street, Cambridge, MA 02138}
\email{dsasselov@cfa.harvard.edu}

\altaffiltext{1}{corresponding author}

\begin{abstract}

The recent discovery of super-Earths (masses $\leq$ 10 $M_{\oplus}$) has initiated a discussion about
conditions for habitable worlds. Among these is the mode of convection,
which influences a planet's thermal evolution and surface conditions.
On Earth, plate tectonics has been proposed as a necessary condition
for life. Here we show, that super-Earths will also have plate tectonics.
We demonstrate that as planetary mass increases, the shear stress
available to overcome resistance to plate motion increases while the
plate thickness decreases, thereby enhancing plate weakness. These effects
contribute favorably to the subduction of the lithosphere, an essential
component of plate tectonics. Moreover, uncertainties in achieving
plate tectonics in the one earth-mass regime disappear as mass increases:
super-Earths, even if dry, will exhibit plate tectonic
behaviour.
\end{abstract}

\keywords{planetary systems --- planets and satellites --- Earth}

\section{Introduction}

Until recently, Earth was the largest terrestrial object known to
exist. However, five super-Earth planets (a class defined as having
a mass between 1-10 $M_{\oplus}$- earth-masses) have been detected
in the last few years \citep{Rivera_et_al:2005,OGLE-5.5:2006,Lovis_et_al:2006,Udry_et_al:2007}.
The five planets have masses in the 5-10 $M_{\oplus}$ range, but we do
not have information on their sizes and cannot be sure if these are
really rocky terrestrial planets. However, their discovery provides
some evidence that super-Earths might be common and it is only a matter
of chance that our Solar System has none. Some of these planets might
be in the 'habitable zone', where the radiation from the star allows
for the presence of liquid water, but only their thermal and chemical
evolution will determine if they are, in fact, habitable. In turn,
their thermal evolution and surface conditions depend on and affect
their tectonic regime. Currently, Earth is the only planet where plate
tectonics is active. Furthermore, this mode of convection has dominated
our planet's geological history, is associated with geochemical cycles
and thus, has been proposed as a required mechanism for life on Earth
\citep{Walker_et_al:1981}. Here we address whether or not super-Earths
are likely to have plate tectonics or be in a stagnant lid convection
like Mercury and Mars.

\section{Analysis}

Plate tectonics is a complicated process that primarily requires lithospheric failure, deformation and subduction. 
For this, convective stresses of the system have to be large enough to overcome lithospheric resistance to deformation 
and the plate has to reach negative buoyancy to drive subduction. The convective stresses are a function
of the underlying flow field and the viscosity of the fluid, whereas
the plate's strength depends on the elastic thickness and mechanical properties. 
Our contribution is based on identifying how the lithospheric or plate
thickness ($\delta$) and convective stresses ($\tau$, $\sigma$)
depend on planetary mass (M). We use detailed models of the internal structure
of massive terrestrial planets \citep{Valencia_et_al:2006, Valencia_ternary} and parameterized
convection models to show that the condition for deformation and subsequent subduction is easily
met by planets more massive than Earth. 
A second condition necessary for subduction is that the plate cools enough to develop negative buoyancy at subduction zones \citep{Sleep:1992, Davies:1993}.
It is unclear as to how important this requirement is with recent elaborated petrological models \citep{Hynes:2005, Afonso_et_al:2007} suggesting that on Earth, the plate's mean density is never larger than the underlying mantle's. Nontheless, we discuss the conditions for the development of negative buoyancy on super-Earths.

\subsection{Parametric Convection}

The thickness of the lithosphere or boundary layer ($\delta$) depends
on the Rayleigh number $(Ra$) -- a parameter governing convection.
A variety of models of boundary layer convection \citep{O'Connell-Hager-1980,Turcotte_Schubert:fluid_mechanics}
lead to $\delta\sim D\left(Ra/Ra_{c}\right)^{s}$, with $s=-1/4$,
and $Ra_{c}$ the critical Rayleigh number above which a fluid starts
convecting. This results in $\delta$ being independent of the depth
$D$ of the convecting layer.

The exponential relationship between $\delta$ and $Ra$ has been
extensively addressed as different flow details, geometries, turbulence,
and other complicating effects are taken into account. In particular,
\citet{Conrad_Hager:1999-GRL} suggested that $s\sim0$ where viscous
dissipation within the bending subducting slab is larger than the
dissipation within the shearing mantle, which is the case for thick
strong plates \citep{Conrad_Hager:1999-JGR}. However, since we show
below that the plates for super-Earths are thin, the case where $s\sim0$
can be ruled out. Other values of $s$ will not change the qualitative
results obtained here.

The deviatoric horizontal normal stress ($\sigma$) responsible for
causing failure on the plate is (to first order) balanced by the shear
stress ($\tau$) applied over the base of the plate. Thus, by a simple
force balance $\sigma=\tau$ $L/ \delta $, where $L$ is the plate
length, calculated from the time it takes for the
plate to cool conductively and grow to a thickness $\delta$,
$L\sim \delta^{2}$ $u / \kappa $, where $u$ is the convective velocity.
The shear stress depends on the viscosity $\eta$ and the velocity
field of the fluid. In the most simple case this dependence will be
linear so that the stress under the plate is $\tau\sim\eta$ $u/D$.
The faster the flow, the larger the stress. In general these are competing
effects: viscosity decreases while velocities increase with higher
temperatures (T). The relationship for velocity 
is $u\sim \kappa/D$ $Ra^{1/2}$ \citep{Turcotte_Schubert:fluid_mechanics}.
A super-Earth can be expected to have larger velocities
due to larger Rayleigh numbers. Thus, we need to investigate the extent
to which the viscosity effect reduces the stress to predict whether
or not plate failure will occur. We do so by considering two cases for viscosity: an isoviscous
case ($\eta=\eta_{0}=10^{21}$Pa$\cdot$s -- Earth's nominal value),
and a T dependent viscosity case ($\eta=\eta_{0}(T/T_{0})^{-30}$
-- \citet{Davies:1980}).

\subsection{Structure and Convection}

We calculate the internal structure of terrestrial planets with masses
in the range of $1-10M_{\oplus}$ \citep{Valencia_et_al:2006} and
determine the mantle thickness, density and gravity of each planet.
We find that a power law relationship can adequately express the dependence
of these parameters on M. For planetary mantles, the Rayleigh number
depends on the mantle density $\left(\rho=\rho{}_{\oplus}(M/M_{\oplus})^{0.20}\right)$,
gravity $\left(g=g_{\oplus}(M/M_{\oplus})^{1/2}\right)$, mantle thickness
$\left(D=D{}_{\oplus}(M/M_{\oplus})^{0.28}\right)$ and heat flux
($q$), which all depend on the mass of the planet. $Ra$ also depends
on material properties: thermal expansivity ($\alpha$), thermal diffusivity
($\kappa$), thermal conductivity ($k$) and viscosity ($\eta)$.
The latter is a strong function of T. Therefore the Rayleigh
number $Ra=\alpha\rho gD^{4}q / \kappa k\eta$, depends indirectly
on M and T. It increases for super-Earths
as the size of their mantle and interior heat increases. To account
for the heat flux as a function of M we scale radioactive heat
production and indirectly consider secular cooling. It is straightforward
to scale the radioactive production with M for similar bulk compositions.
Secular cooling, on the other hand, can only be completely assessed
by considering the full thermal evolution. More massive planets would
have larger initial gravitational energy budgets to dissipate over
time; we assume for simplicity that the heat to be lost to secular
cooling is proportional to M. Convective velocities also scale
roughly with M ($u\sim M^{1.19}$-- see Table 1), so
that the proportional rate of cooling should be roughly independent
of M. Even though the relative contributions of radioactive
heating and secular cooling to the total heat flow will depend on
the evolution of the planet, we will assume that heat flow scales
proportionately with M, especially because we expect the radioactive
production to be dominant. The internal structure model allows us
to calculate Ra, $\delta$ and T beneath the plate for planets between 1-10 $M_{\oplus}$.
Convergence of these parameters determines the structure and final
radius of each super-Earth planet.

\section{Results}

Figure 1 shows the dependence of plate thickness (blue) and horizontal normal
stress (green) as a function of M for the T-dependent-viscosity case and Earth-like radioactive heat sources (solid). An additional case with reduced heat sources (dashed) is discussed below.  The trends
between these two scenarios and scaling exponents are similar, although their $1M_{\oplus}$ values are different.  

As expected, the plate thickness decreases as the M of the planet increases. This is because a more vigorous convective interior can transport heat more efficiently
to the surface and sustain a higher surface heat flux. In addition,
shear stress underneath the plate increases proportionately with M (Fig. 2),
and thus, the deviatoric horizontal normal stress also increases.
This means that the viscosity reduction effect on stress is very small
compared to the velocity effect. This is the case because the drop
in T within the plate and hence local viscosity are
nearly independent of M as indicated by our calculations for the more realistic T-dependent-viscosity case
 (Fig. 2). For similar surface temperature, the increase in temperature
at the base of the lithosphere between Earth and a super-Earth differs
at most by only 5\%. We think the reason for this surprising result
is a very efficient negative feedback induced by viscosity. If T increases, the viscosity decreases and heat flux increases,
cooling the planet and reducing T. We ignore pressure (P)
effects on viscosity because we expect them to be small under thin
plates and because we find that P at the base
of the plate is roughly the same for all super-Earths.

Figure 2 shows the dependence of all the relevant convective parameters
with planetary mass normalized to the Earth's values for the T-dependent case. The results can be adequately fitted to a power law
such as $\delta=\delta_{\oplus}(M/M_{\oplus})^{\beta}$ where the
exponent $\beta$ shows the dependence on M. In a vigorously
convective interior (high $Ra$) the velocities are expected to be
large and this is evident in Fig. 2. The relative
size of the convection cells measured as the length of the plate with
respect to its radius ($L/R$) increases slightly, from a
0.29 for a $1M_{\oplus}$ planet to a 0.30 ratio for a $10M_{\oplus}$
planet. The maximum plate age, related to the convective overturn,
in a super-Earth is much less (see Table 1 for $\beta=-0.91$).

\subsection{Heat Flow}
In addition to a bulk silicate Earth composition for radiactive sources, we also consider planets with no $^{40}K$ content.  As a result of potassium's volatile  nature, Earth is depleted in this element \citep{McDonough_Sun:1995}. It is therefore possible for other planets to also have non-chondritic K concentrations. U/O and Th/O ratios can be expected to be fairly constant for planets forming around stars with a relatively recent supernova event (within 10 Gy). K contributes less than 1/3 of total heat from radioactive sources at the present time on Earth. Within this context, we consider a pesimistic scenario for a super-Earth with only 1/3 of its chondritic heat flow ($Q$) and no secular cooling contribution.

This end-member case ($Q/3$) shows comparatively thicker plates (Fig. 1), with 88 km versus 43 km in thickness for a $1M_{\oplus}$ planet. Additionally, the driving force is in the order of 1-6 MPa which is lower than the Earth's value of $\sim$10 MPa.  This is an extreme case where subduction would be difficult but still possible for the most massive super-Earths if other factors, such as the presence of hydrated minerals is invoked to reduce the strength of the plate. For each planetary mass, there is a threshold in heat flow below which the convective driving force would not be sufficient to maintain subduction and plate tectonics would cease. Nontheless, most of the super-Earths that will be detected in the next few years will likely fall in a regime above the $Q/3$ scenario.

\subsection{Lithospheric Buoyancy}
On Earth, the oceanic lithosphere is comprised of 7 km of basaltic crust (chemically buoyant), overlaying lithospheric mantle. We perform a simple buoyancy calculation at subduction ages by considering the thermal contraction from the ridge. We determine maximum crustal thicknesses that still allow the lithospheric mean density to be larger than the underlying mantle's. We find this value to be 13\% the plate thickness at subduction zones (twice the average thickness - $\delta$). The crustal thickness depends on the extent of melting under ridges which depends on the P and T conditions. As shown above, T at ridges and under plates varies little with M, so that with larger gravity values, melting occurs at shallower depths in super-Earths. A simple calculation shows that the crust thins with increasing planetary M, and while it becomes a larger component of the also thinning boundary layers, its thickness stays below the threshold allowing negative buoyancy. This effect stands in contrast to the young Earth that had a higher potential T but same P profile \citep{Nisbet_Fowler:1983, Davies:1993}. To precisely assess the density of the plate detailed petrological models like those of \citet{Hynes:2005} and \citet{Afonso_et_al:2007} are needed. It is important to point out that their models do not predict negative buoyancy on Earth when the plate's density is compared to sublithospheric density, arguing that perhaps this condition is not critical for the initiation of subduction. In addition, \citet{Becker_et_al:1999} showed that in compressive regions, the cold plate can thicken to a point that negative buoyancy is inevitable and subduction follows. If negative buoyancy is indeed a necessary condition, our simple calculations show that super-Earths can satisfy it.

\section{Summary and Implications}
In summary, convection is more vigourous in massive terrestrial planets,
making their lithospheres thinner and therefore reducing lithospheric
strength. Furthermore, they achieve larger stresses owing primarily
to larger velocities and therefore can more easily overcome the lithospheric
resistance to deformation. Plates may reach negative buoyancy on super-Earths despite their relative younger ages. This scenario is suitable for the failure
of the plate and subsequent subduction, which is a necessary step
for plate tectonics. Given that Earth's convective state leads to plate tectonics, 
the more favorable conditions experienced by super-Earths will inevitably lead to plate tectonics.
Furthermore, planets of similar mass should have the same potential to exhibit
plate tectonics. Conversely, this physics can help explain why small
planets like Mars, Mercury and the Moon do not exhibit plate tectonics.

\subsection{Role of Water}
Venus is only slightly smaller than Earth and does not exhibit plate
tectonics, although some authors \citep{Turcotte:1993,Jellenik_water:2005J}
have suggested it may have in the past. This observation indicates
that the $\sim$1 earth-mass case falls within a zone
of transition between `hard' stagnant lid and mobile plate regimes.
In this case, characteristics other than M may be important to
the dynamics of the lithosphere. For example, the high surface temperature
of Venus might lead to a weak, highly deformable boundary layer that
would not support the coherent plate-like behaviour that characterizes
oceanic plates on Earth. Moreover, plate strength is relatively large
compared to the mantle driving force in the one earth-mass case; yield
stresses are on the order of 1-5 GPa for olivine (the representative
upper mantle mineral) \citep{Chen_et_al:1998}, whereas our calculations,
in agreement with more detailed models \citep{Becker_Oconnell:2001},
suggest an underlying driving force of only 10 MPa. Since slip can
occur on pre-existing faults at stress values of a few MPa, the existence
of plate tectonics in the one earth-mass regime may thus depend crucially
on the conditions required to initiate subduction. The presence of
water is one possible mechanism to reduce the yield strength of a
plate and friction on faults. Experiments show that water reduces
the yield strength of olivine by 62\% when raising the temperature
from 25 to 400$^{\circ}$C at 10 GPa, compared to a drop of 39\% in
dry olivine \citep{Chen_et_al:1998}. Hence, the hydration level of
Venus' mantle, which is 1-2 orders of magnitude lower than on Earth \citep{Zolotov_et_al:1997},
may make it very difficult for convective forces within this planet
to overcome plate resistance.

For larger planets, super-Earths, these issues become less relevant.
A wet super-Earth will clearly have enough driving force to sustain
subduction.  But, more importantly, the consequences for initiating
  subduction associated with the hydration of a one earth-mass planet
  (i.e., a reduction of the yield strength by half) would be similar to
  a doubling of the mass of the planet (Fig. 2). That is to say, both 
scenarios would be as likely to initiate and maintain subduction.

\subsection{Atmospheric Observables}
The difference between a Super-Earth with active plate tectonics and one with stagnant lid is in the access of upper mantle material and gasses to the atmosphere. The first case allows several global geochemical cycles to operate, like the CO$_2$ and SO$_2$ ones. For example, cases in our Solar System comprise:  Earth with a CO$_2$ cycle and possibly early Mars with a SO$_2$ cycle \citep{Halevy_et_al:2007}. Earth has had stable modest levels of atmospheric CO$_2$ (between 160-7000 ppm -- \citet{Royer_et_al:2001}) in the last 0.5 Gy whereas Venus’ levels stand today at 96\%. A planet with plate tectonism and a carbonate rock reservoir has an efficient built-in cycle that stabilizes climate at temperatures within the liquid water regime \citep{Kasting:1996}. A super-Earth that has plate tectonics and weathering capabilities can be expected to have CO$_2$ atmospheric concentrations that would yield temperatures around liquid water. Therefore, evidence against the presence of plate tectonics on an exoplanet would be the detection of high values of CO$_2$ for the age of the star, type of star and orbital distance. An SO$_2$ based atmosphere is also possible and the same reasoning would apply, since the sulfur cycle operates analogously to the carbon cycle. But obviously, more theoretical research is necessary to model the details and predict the right observable signatures.

In conclusion, we show here that as mass increases, the process of subduction, and hence plate tectonics, becomes easier. Therefore, massive super-Earths will very likely
exhibit plate tectonics. In the future with TPF by NASA and Darwin by ESA it might be possible to use spectroscopy to identify atmospheric signatures suggesting plate tectonism on these objects. This class of planets offers the possibility of finding Earth analogs and, in particular, make attractive targets in the search for habitable planets.

\section{Acknowledgements}
DV acknowledges support from the Harvard Origins of Life Initiative. This work was in part funded by NSF grant EAR=0440017.

\clearpage

\begin{figure} [h] {\bf Fig. 1}
\plotone{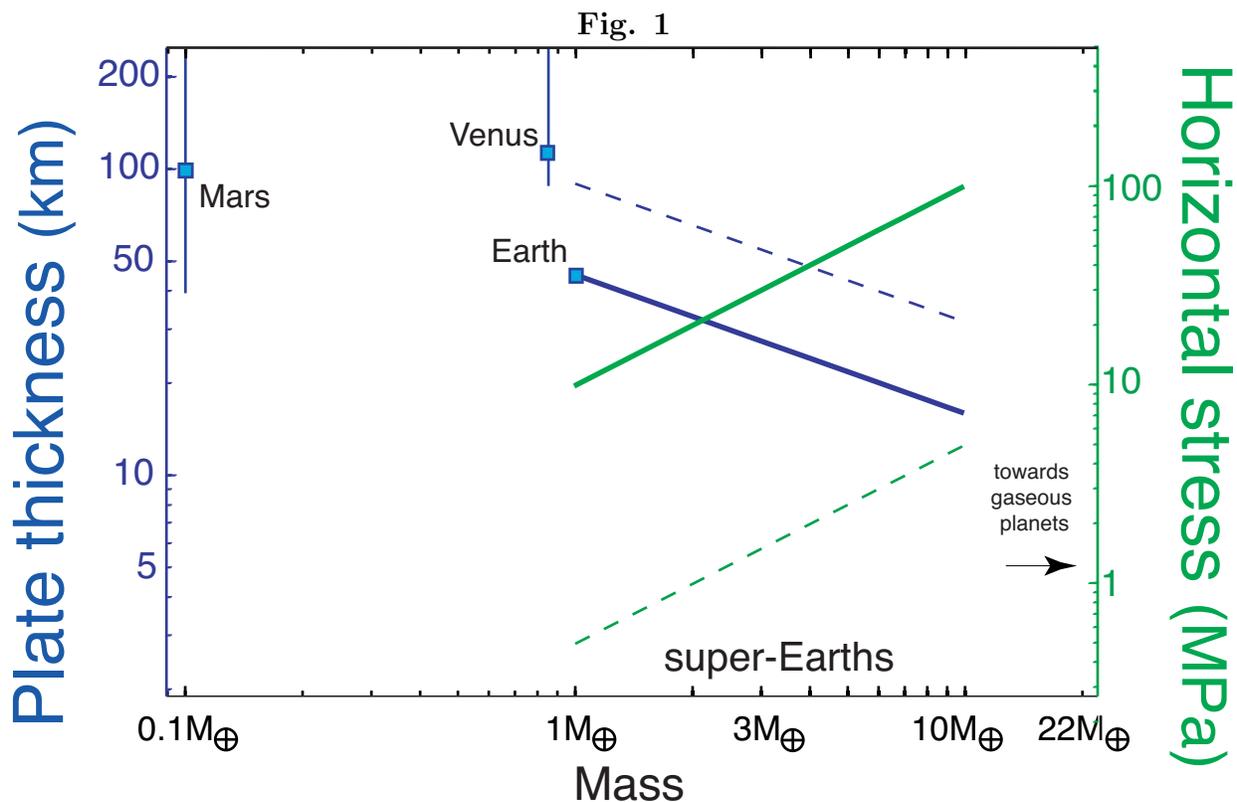}
\figcaption{ Plate thickness and stress for super-Earths. The dependence of plate
thickness (blue and left axes) and horizontal normal stress (green and right
axes) on mass are shown for a BSE composition (solid) and a reduced heat content case (dashed). Both cases have a T-dependent viscosity treatment. As mass increases, the stress
available to overcome plate resistance increases, while plate thickness
decreases, enhancing plate weakness. Plate thickness for Mars and Venus
is shown with their uncertainty (from \citep{Zuber:2000,Nimmo_McKenzie_Venus:1997}).
The Earth's plate thickness is a global average. Planets with masses
larger than 10$M_{\oplus}$ can start to retain gas during formation
\citep{Ida:2004}.}
\end{figure}
\clearpage

\begin{figure}[h] {\bf Fig. 2}
\plotone{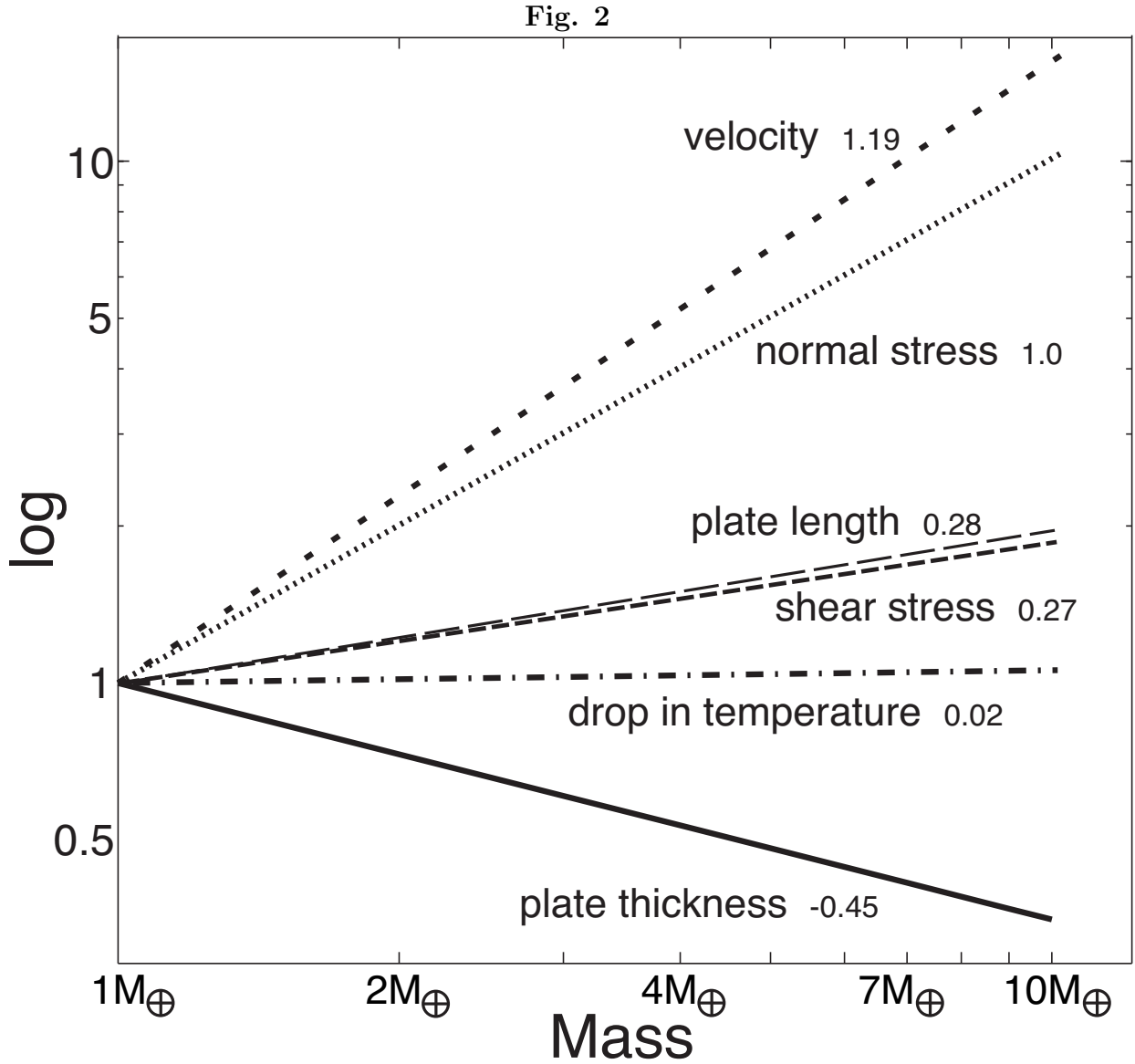}
\figcaption{Convective parameters scaling with planetary mass. Plate thickness (solid line), drop in temperature within the plate (dash-dotted line), shear stress (medium dashed line), plate length (long dashed line), normal stress on plate boundaries (dotted line) and convective velocity (short dashed line) depend on the mass of the planet. This dependence can be fitted to a power law with corresponding exponents shown as numbers. The values are normalized for the Earth - see Table 1.}
\end{figure}

\clearpage

\pagebreak

\begin{deluxetable}{cccc}
\tablewidth{0pt}
\tablecolumns{4}
\tablecaption{Convective parameters' power law dependence on mass}
\tablehead{
\colhead{Parameter} & \colhead{Earth} & \colhead{$\eta=\eta(T)$} & \colhead{Isoviscous} }
\startdata

plate thickness $(\delta)$& 43 km & -0.45& -0.29\\
drop in temperature $(\Delta T)$& 1277 $^{\circ}$C& 0.02& 0.18\\
plate's velocity $(u)$& 3 cm/y & 1.19 & 0.87\\
shear stress $(\tau)$& 0.3 MPa & 0.27 & 0.58\\
normal stress $(\sigma)$& 10 MPa& 1.00 & 1.16\\
plate length $(L)$& 1800 km& 0.28& 0.29\\
convective time $(t)$& 70 myr& -0.91 & -0.58\\

\enddata
\tablecomments{The second column displays Earth's convective parameters. 
The only assumed value is the plate velocity. The third and fourth column show the value
for the exponent $\beta$ in the power law relationship -- ie. $\delta=\delta_{\oplus}(M/M_{\oplus})^{\beta}$ -- 
for a BSE composition with a  T-dependent viscosity and isoviscous cases respectively.}

\end{deluxetable}


\end{document}